\documentclass[twocolumn,showpacs,aps]{revtex4}
\usepackage{graphics}
\newcommand{\rb}{{\mathbf r}}
\newcommand{\eb}{{\mathbf e}}

\newcommand{\Rb}{{\mathbf R}}
\newcommand{\Ri}{{{\mathbf R}_i}}
\newcommand{\Ab}{{\mathbf A}}
\newcommand{\Bb}{{\mathbf B}}
\newcommand{\Cb}{{\mathbf C}}
\newcommand{\Ec}{{\mathcal E}}
\newcommand{\mPhi}{{\mathit\Phi}}
\newcommand{\mOmega}{{\mathit\Omega}}

\newcommand{\sump}{\mathop{{\sum}'\!\!\!}}
\begin{document}
\title{Translational invariance of Coulomb series and symmetric
potentials in crystals}
\author{Eugene V. Kholopov}
\email{kholopov@casper.che.nsk.su}
\affiliation{Institute of Inorganic Chemistry of the Siberian Branch of
the Russian Academy of Sciences, 630090 Novosibirsk, Russia}
\date{}
\begin{abstract}
It is shown that Coulomb series are to be considered within a special mode
of summation so as to describe bulk properties of crystals. The
translational invariance is then an explicit integral property of Coulomb
series that is tantamount to the effect of invariant periodic boundary
conditions discussed earlier. Absolute bulk potentials with zero mean
value are then substantiated as a unique solution in the general case of
triclinic lattices. An invariant treatment of the bulk Coulomb energy
follows therefrom. The potential symmetry is verified for simple
point-charge lattices and is connected with the centre of gravity of the
potential field that is relevant to non-local charges as well.
\end{abstract}
\maketitle
\section{Introduction}\label{Sec1}
Sums of Coulomb potentials over lattices \cite{Glas80} are basic for
describing the cohesive energy of solid state \cite{Sher32,Nijb57,Wadd59,%
Tosi64,Tref87,Ihm988,Bagn91}. Such sums also determine local potentials
responsible for the electronegativity and charge transfer in crystals
\cite{Tref87,Levi66,Goo69b,Wang80,Sat81b,Spac81,Gene87,Cole97,Koha98} and
so specify the electronic band structure \cite{Ihm988,Bagn91,Levi66,%
Spac81,Call58,Rudg69,Euwe75,Harr75,Dove83,Angy87,Scha92,King93,Fuch94,%
Zhan94,Alat99}.

The problem which still exists is to determine bulk Coulomb series so that
the result of summation in all the above events be unique in every
particular case. The potential symmetry is then evident at least in the
simplest point-charge lattices of AB type \cite{Levi66,Roy954,Birm55,%
Goo69a,Cala76,Xiao92}. Therefore, the bulk Coulomb energy can be specified
by particular values of the absolute local potentials therein
\cite{Nijb57,Roy954,Goo69a,Ewal21,Evje32,Hoje38,Gurn53,Emer55,Harr70,%
Tayl87,Argy92}. However, different modes of summation with various types of
parametrization either in the direct space \cite{Roy954,Cala76,Evje32,%
Made18,Fran50,Dahl65,Hajj79,Mass82,Coke83,Goni91,Fise92,Wolf92,Vito95} or
with making use of the reciprocal one \cite{Nijb57,Ewal21,Harr70,Mass82,%
Emer23,Bert52,Shol67,Sugi84,CraB87,Luty95} lead to controversial results.
This fact is due to the conditional character of convergence of infinite
Coulomb series \cite{Glas80,Epst03,Borw85}.

To be unique, the bulk solution of interest must, certainly, be
independent of the optional parameters of the mode of summation
\cite{Roy954,Dahl65,Hajj79,Mass82,Redl75}. As a hint about it, surface
effects are to be irrelevant to bulk properties \cite{Bagn91,Harr75,%
Roy954,Cala76,Ewal21,Hajj79,Coke83,Wolf92,Redl75,Coog67,HeyS81,Mart97}.
All the statistical treatment of solids is based on this statement
\cite{Born54,Zima60,Grif68}. Zero mean bulk potential is declared
therefrom regardless of the summation scheme \cite{Bagn91,Harr75,Ewal21,%
Gurn53,Harr70,Dahl65,Luty95,Redl75,HeyS81,Jain80,Herz85}. But it is
neither the case of parent infinite series \cite{Heye81,Klei81} nor the
case of periodic boundary conditions \cite{Born54,Born12} while they may
be modified by a constant potential as an available periodic solution of
Laplace's equation \cite{Rudg69,Zhan94,Dahl65,Redl75}.

As found \cite{Khol01,Khol02}, the solution of interest arises under
periodic boundary conditions invariant to the definition of the unit cell
relative to the crystal structure, which generalize the translational
invariance based on periodic replications of a certain basic unit cell
\cite{Bagn91,Harr75,Redl75,Leeu80,Mako95,Schu99}. This result
substantiates Ewald's proposal \cite{Ewal21} to exclude the uniform
potential component \cite{Redl75,Coog67,Mako95} as well as the energy
effect of surface polarization \cite{Leeu80,Mako95,Smit81,Yoko85} from the
results of summation that was not convincing for a long time
\cite{Bagn91,Euwe75,Harr75,Dahl65,Mako95,Wett80,Frec85,Oliv86,Deem90,%
Keef94}.

Moreover, though the energy value for a neutral unit cell in the bulk is
formally independent of any conceivable change in the potential origin
\cite{Bagn91,HeyS81,Mart97,Mako95,IhmC80,Rest90}, the bulk Coulomb energy
and absolute bulk potentials appear to be functionally connected. The fact
that electrostatic quantities in the bulk are independent of real surfaces
implies that the configuration of any surface is to be consistent with the
bulk state \cite{Tosi64,Levi66,Angy87,Wolf92,HeyS81,Smit81,Task79,LeeC80,%
Wats82,Kokk95,Plih98}. In particular it means that the surface conditions
have to reproduce the bulk state in question \cite{Made18,HeyS81,Klei81,%
Mack57}.

It is important that the same problem is relevant to the definition of
direct bulk Coulomb series as such. Indeed, pristine Coulomb series,
albeit expressed in terms of periodic charge distributions, do not contain
the periodicity as a property of summation in an explicit fashion
\cite{Ewal21}. In the present paper we deduce that invariant periodic
boundary conditions actually imply a special mode of summation as a
continuous procedure with piecewise smooth intermediate boundary surface
and with the further averaging of the oscillating result over the period
specified by lattice translations. As shown, apart from the convergence
conditions, an additional one should be imposed on the charge distribution
per unit cell so as to represent Coulomb series in a conventional form.
As an example, the symmetric potentials in cubic crystals are readily
calculated in an effective manner. The topological nature of the potential
symmetry is discussed.

\section{Ambiguity about absolutely convergent Coulomb series}
\label{Sec2}
If $\rho(\rb)$ is a certain charge distribution  attributed to a unit
cell, then the direct Coulomb sum associated with $\rho(\rb)$ and
describing the electrostatic potential at the point $\rb$ can be written
as
\begin{equation}\label{Aq1}
U_{\mathrm{Cd}}(\rb)=\sump_i\int_V\frac{\rho(\rb')\:d\rb'}{|\rb-\rb'
-\Ri|},
\end{equation}
where $i$ runs over the Bravais lattice composed of the unit-cell origins
$\Ri$, the integral is over the volume $V$ occupied by $\rho(\rb')$, which
can spread beyond the unit-cell volume \cite{Tref87,Dove83,Scha92,Zhan94,%
Alat99,Birm55,Goni91}, the prime on the summation sign implies missing the
contribution of any point charge if it happens at $\rb$. For the absolute
convergence of (\ref{Aq1}), $\rho(\rb)$ is supposed to obey the following
constraints \cite{Harr75,Cala76,Evje32,Dahl65,Fise92,Emer23,Redl75,%
Coog67,Herz85,Mako95,Hall79}:
\begin{eqnarray}
&&{\displaystyle\int_V}\rho(\rb)\:d\rb=0 ,\label{Aq2}\\
&&M_\mu\equiv{\displaystyle\int_V}r_\mu\rho(\rb)\:d\rb=0 , \label{Aq3}\\
&&G_{\mu\nu}\equiv{\displaystyle\int_V}r_\mu r_\nu\rho(\rb)\:d\rb=0
\quad\mbox{at}\quad\mu\neq\nu ,\label{Aq4}\\
&&G_{xx}=G_{yy}=G_{zz}=H ,\label{Aq5}
\end{eqnarray}
where $r_\mu$ are Cartesian co-ordinates of $\rb$, $H$ is a certain
constant. Condition (\ref{Aq2}) ensuring the local neutrality is
conventional. Relations (\ref{Aq3})--(\ref{Aq5}) can be maintained by
symmetry \cite{Bagn91,Call58,Evje32,Dahl65,Redl75,Coog67,Jain80}. If they
do not hold for any initial $\rho^{\mathrm{ini}}(\rb)$ \cite{Bagn91,%
Angy87,King93,Evje32,Coog67,Frec85,Rest90,Hall79,Sat81a}, $\rho^{\mathrm
{ini}}(\rb)$ can always be modified properly \cite{Khol01,Khol02}. The
implication is that the electrical neutrality is the only fundamental
constraint in crystals \cite{Bagn91,Harr75,Ewal21,Hoje38,Emer23,CraB87,%
Heye81,Smit81,Oliv86,Deem90,CraD87}, but a rather arbitrary charge
distribution $\rho^{\mathrm{ini}}(\rb)$ transformed into $\rho(\rb)$
driven by (\ref{Aq2})--(\ref{Aq5}) with an optional $H$ may be adopted for
the purpose of the lattice summation specified as a definite procedure in
(\ref{Aq1}) \cite{Dove83,Roy954,Ewal21,Coke83,Emer23}.

The residual effect of $H$ is evident from the mean potential value
$\bar{U}_{\mathrm{Cd}}$ associated with (\ref{Aq1}) and it is independent
of the order of summation due to (\ref{Aq2})--(\ref{Aq5}). Following Bethe
\cite{Beth28}, we have
\begin{equation}\label{Aq6}
\bar{U}_{\mathrm{Cd}}=-\frac{2\pi}{3v}\int r^2\rho(\rb)\:d\rb
=-\frac{2\pi H}{v}\; ,
\end{equation}
where $v$ is the volume of the unit cell, the left equality of (\ref{Aq6})
is typical \cite{Bagn91,Call58,Harr75,Klei81,Mako95,Keef94,Rest90,Beck90}
and is further converted with the help of (\ref{Aq5}). We see that the
value of $\bar{U}_{\mathrm{Cd}}$ turns out to be optional through $H$.
This inference can be naturally extended to the potential
$U_{\mathrm{Cd}}(\rb)$ as such \cite{Harr75,Klei81}. However, it is also
clear that the effect of $H$ is artificial and should be removed.

Although the case of $H=0$ is of great importance, as will be clear later
on, it does not exhaust the situation at hand. To gain insight into this
problem, we turn to the limiting behaviour of (\ref{Aq1}) upon the direct
summation over closed shells of one unit cell thickness, enveloping a
reference unit cell in a consecutive fashion. According to
(\ref{Aq2})--(\ref{Aq5}), such an order of summation is not more demanded
by the arguments of convergence \cite{Harr75,Zhan94,Roy954,Ewal21,Evje32,%
Hoje38,Tayl87,Coke83,Wolf92,Emer23,Bert52,CraB87,Borw85,Coog67,Leeu80,%
Smit81,Bhow88}, but allows for the equal uniformity along each
crystallographic axis \cite{Dove83,Emer23,Redl75,HeyS81,Klei81}. The
absolute convergence of (\ref{Aq1}) implies that the contribution of
remote shells with numbers $m\geq m_{\mathrm c}\gg1$ is negligible:
\begin{equation}\label{Aq7}
\lim\limits_{m_{\mathrm c}\to\infty}\sum_{m=m_{\mathrm c}}^{\infty}
\sum\limits_{i_m}\int_V\frac{\rho(\rb)\:d\rb}{|\Rb_{i_m}+\rb|}=0 ,
\end{equation}
where $i_m$ runs over the unit cells belonging to the $m$th shell.
Relation (\ref{Aq7}) persists as long as remote unit cells are added as
permanent bricks described by $\rho(\rb)$ \cite{Bagn91,Harr75,Angy87,%
Evje32,Hajj79,Redl75,Coog67,HeyS81,Heye81,Klei81,Keef94,LeeC80,Hall79}.
Since individual features of such bricks have to be lost in the crystal
interior \cite{Harr75,Evje32,Hajj79,Redl75}, the fortuitous contribution
of $H$ to the lattice sum is patently generated by outer shells of
summation \cite{Harr75,Redl75,Keef94} and so must be eliminated as
irrelevant to bulk properties now.

\section{Continuity of summation and invariant periodicity}\label{Sec3}
\begin{figure}[t]
\resizebox{0.9\hsize}{!}{\includegraphics{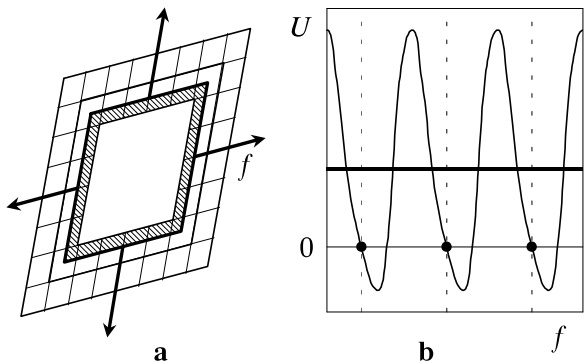}}
\caption{$\!\!$Summation over the shells as a uniform continuous procedure:
(a) Remote unit cells, small regular parallelograms, are united into the
shells bordered by heavier lines, with the hatched area of an incomplete
outermost shell truncated by the bounding box depicted by the heaviest
lines. The bounding-box interior contributing to the lattice sum
enlarges uniformly upon increasing the evolution parameter $f$. (b)
Oscillations of the limiting value of a certain inner potential $U$ as a
function of $f$ are expected relative to its magnitude specified by
complete outermost shells. This initial magnitude marked by the filled
circles is treated as an origin, for convenience. Averaging over
oscillation periods shown by vertical dotted lines converts the
oscillating $U$ into an invariant value depicted by the heavy horizontal
line, of which position describes the potential shift of interest.}
\label{Fig1}
\end{figure}

Inasmuch as (\ref{Aq7}) holds due to the discrete character of summation
over shells there, it is reasonable to modify this type of summation so as
to remove that discreteness, but without destruction of the charges taken
into account \cite{Harr75,Argy92,Jain80,CraD87}. To this end, we conceive
that our crystal is enclosed in an imaginary large bounding box homothetic
to the unit cell, with a reference unit cell at its centre, and the
summation region is restricted by the box volume. The summation over
shells is then treated as a continuous procedure of enlarging this box. In
a triclinic lattice with the lattice parameters $a$, $b$ and $c$ along
unit vectors $\eb_a$, $\eb_b$ and $\eb_c$, the evolution of the bounding
box can be described by the parameters $A$, $B$ and $C$ corresponding to
$\eb_a$, $\eb_b$ and $\eb_c$ and increasing linearly in an evolution
parameter $f$ at
\begin{equation}\label{Bq1}
\frac{A}{a}=\frac{B}{b}=\frac{C}{c}\gg1 ,
\end{equation}
as shown schematically in Fig. \ref{Fig1}(a).

The event of $\rho(\rb)$ arranged within the unit-cell parallelepiped is
first considered. The limiting procedure in (\ref{Aq7}) can then be
readily associated with the bounding-box evolution if we imagine that the
$m_{\mathrm c}$th shell is truncated by the bounding box at a given
moment. It implies that the inner part of this shell gives an additive
contribution to the maternal sum over $m<m_{\mathrm c}$ and this shell is
removed from (\ref{Aq7}) and replaced by the next one when its outer
boundary is achieved by the bounding box. The periodicity of the latter
events upon the evolution in question reflects the translational
invariance of (\ref{Aq7}) accompanied by some periodic change in the value
of the resulting maternal sum, as sketched in Fig. \ref{Fig1}(b). This
limiting effect can naturally be analysed within a single period of
oscillations at $0\leq f\leq1$, so that $f$ becomes a measure of the
incorporated share of the $m_{\mathrm c}$th shell.

Note that now the expected result depends on our summation mode exhibiting
the translational invariance as an integral property of Coulomb series
\cite{Ewal21}. To make the above oscillations surely finite and minimal,
all the faces and edges of the bounding box are assumed to be so charged
as to retrieve local constraints (\ref{Aq2}) and (\ref{Aq3}) for the
truncated unit cells at every moment of the bounding-box evolution. Then
the potential effect of the remote charged edges of the truncated
$m_{\mathrm c}$th shell modified by charges on the corresponding edges of
the bounding box vanishes because it is specified by the mean charge
densities \cite{Roy954,Redl75,Coog67,HeyS81,Task79}, which are zero by
definition.

To discuss the contribution of the modified $m_{\mathrm c}$th shell faces
to (\ref{Aq7}), we focus on the $\Ab$th face disposed in the $\eb_a$
direction, with $\Ri$ describing the unit cells on this face. Here we are
interested in the potential contribution at the centre of the bounding
box. In the expansion of the integrand in (\ref{Aq7}) in powers of
$r/R_i$, the ratio of moduli, only terms linear in the components
$R_i^{\bot\Ab}$ of $\Ri$ along an outward normal to the $\Ab$th face
survive \cite{Coke83,HeyS81,Task79,LeeC80}. Their contribution to
(\ref{Aq7}) can be written as
\begin{eqnarray}\label{Bq2}
\mPhi_{\Ab}(f)&=&\frac{\mOmega^2}{\sin\beta}\int_{-a/2}^{t(f)}
dt\int_{-b/2}^{b/2}dp\int_{-c/2}^{c/2}du\,
\rho(t,p,u)\nonumber\\
&&{}\times\Bigl[t(f)-t\Bigr]\sum\limits_{i\in\{\Ab\}}
\frac{R_i^{\bot\Ab}}{R_i^3}\; ,
\end{eqnarray}
where $t(f)=af-a/2$, $\rb=t\eb_a+p\eb_b+u\eb_c$,
\begin{eqnarray}\label{Bq3}
\mOmega&=&\bigl[1-\cos^2\alpha-\cos^2\beta-\cos^2\gamma\nonumber\\
&&{}+2\cos\alpha\;\cos\beta\;\cos\gamma\bigr]^{1/2}
\end{eqnarray}
at $(\eb_a\eb_b)=\cos\alpha$, $(\eb_b\eb_c)=\cos\beta$ and
$(\eb_c\eb_a)=\cos\gamma$.

The averaging of $\mPhi_{\Ab}(f)$ over the period of oscillations is
readily defined in the form
\begin{equation}\label{Bq4}
\bar{\mPhi}_{\Ab}=\int_0^1\mPhi_{\Ab}(f)\,df
\end{equation}
and is the final logical step of elimination of the summation boundary,
because the boundary which has neither definite structure nor definite
position does not more exist. This case is naturally associated with the
topologically endless lattice. Indeed, the same result is valid for the
contribution $\bar{\mPhi}_{-\Ab}$ of the $-\Ab$th face of the $m_{\mathrm
c}$th shell by symmetry, but their combination may be represented as
\begin{equation}\label{Bq5}
\int_0^1\Bigl[\mPhi_{\Ab}(f)+\mPhi_{-\Ab}(1-f)\Bigr]df
=2\bar{\mPhi}_{\Ab} ,
\end{equation}
where the integrand corresponds to the closure of the lattice in the
$\eb_a$ direction along the $\pm\Ab$th bounding-box faces, in which case
these auxiliary bounding-box faces merge and so neutralize each other by
definition, but periodic boundary conditions arise instead. The averaging
over $f$ in (\ref{Bq5}) then implies the invariance of the periodic
boundary conditions to the optional position of the plane of joining
inside particular unit cells \cite{Khol01,Khol02}.

On determining $\bar{\mPhi}_{\Bb}$ and $\bar{\mPhi}_{\Cb}$ in the same
manner and taking our previous result \cite{Khol01,Khol02} into account,
for the total potential eliminating the boundary effect we obtain
\begin{equation}\label{Bq6}
\mPhi_{\mathrm{top}}=2\left(\bar{\mPhi}_\Ab+\bar{\mPhi}_\Bb
+\bar{\mPhi}_\Cb\right)=\frac{2\pi H}{v}\; .
\end{equation}
Note that issue (\ref{Bq6}) is relevant to the foregoing periodic
boundary conditions, which hold for each basic crystallographic direction
and so are three-dimensional. The duality of the treatment of the
procedure of averaging makes this inference geometrically
noncontradictory.

It is significant that like (\ref{Aq6}), result (\ref{Bq6}) is independent
of the particular features of the crystal symmetry. Owing to the obvious
additivity of the potential effect, result (\ref{Bq6}) can be
\begin{table*}[t]
\caption{Tentative unit-cell point charges $s_j$ in units of $\kappa$ for
each of $n_L$ primitive cubic lattices, with the charge positions
generated from those indicated, in units of $a$, in the parentheses by
virtue of cubic transformations.} \label{Table1}
\begin{ruledtabular}
\begin{tabular}{lcccccccccc}
Crystal&${\displaystyle n_L}$&$\!s_1 (0,\!0,\!0)$&$\!s_2 (\frac{1}{4},\!
\frac{1}{4},\!\frac{1}{4})$&$\!s_3 (\frac{1}{2},\!0,\!0)$&
$\!s_4 (\frac{1}{2},\!\frac{1}{2},\!0)$&
$\!s_5 ($-$\frac{3}{4},\!\frac{1}{4},\!\frac{1}{4})$&
$\!s_6 (\frac{1}{2},\!\frac{1}{2},\!\frac{1}{2})$&$\!s_7 (1,\!0,\!0)$&
$\!s_8(\frac{3}{4},\!\frac{3}{4},\!\frac{3}{4})$&
$\!s_9(\frac{3}{2},\!\frac{3}{2},\!\frac{3}{2})$\vspace{0.5mm}\\
\hline
&&&&&&&&&&\vspace{-2.7mm}\\
CsCl&$1$&$\frac{337}{256}$&---&---&---&---&$-\frac{513}{4096}$&
$-\frac{27}{512}$&---&$\frac{1}{4096}$\vspace{1mm}\\
NaCl$$&$4$&$\frac{32}{23}$&---&$-\frac{7}{46}$&$-\frac{3}{92}$&---&
$-\frac{1}{92}$&---&---&---\vspace{1mm}\\
ZnS&$4$&$\frac{415}{384}$&$\;\;-\frac{3}{16}\,$\footnotemark[1]&---&
$-\frac{1}{144}$&$\;\;-\frac{1}{48}\,$\footnotemark[1]&---&
$\frac{1}{2304}$&---&---\vspace{1mm}\\
Cu$_2$O&$\;\;4\,$\footnotemark[2]&$\frac{17}{8}$&
$\;\;-\frac{3}{8}\,$\footnotemark[3]&
---&$-\frac{3}{32}$&$\;\;-\frac{3}{64}\,$\footnotemark[3]&---&---&
$\;\;\frac{1}{64}\,$\footnotemark[3]&---\vspace{1mm}\\
CaF$_2$&$4$&$\frac{35}{16}$&$-\frac{49}{256}$&---&$-\frac{1}{64}$&
$-\frac{5}{256}$&---&---&---&---\vspace{1mm}\\
BiF$_3$&$4$&$\frac{591}{184}$&$-\frac{1}{4}$&$-\frac{61}{368}$&
$-\frac{13}{736}$&---&$-\frac{1}{1472}$&---&---&---\vspace{0.5mm}
\end{tabular}
\end{ruledtabular}
\footnotetext[1]{These charges are at the proper tetrahedral positions
only.}
\footnotetext[2]{\mbox{The final structure is made up of primitive lattices
rotated about a fourfold symmetry axis.}}
\footnotetext[3]{These charges are on $(1,1,1)$ diagonals only.}%
\vspace{-2pt}
\end{table*}
extended to overlapping $\rho(\rb)$, as well as to more complicated
Bravais lattices of higher symmetry. By induction, the validity of
(\ref{Bq6}) can also be extended to events of $\rho(\rb)$ spreading to
\begin{table*}
\caption{As a result of the direct summation over point-charge cubic
lattices, the absolute bulk potentials $U_{{\mathrm b}j}$, in units of
$\kappa/a$, at the unit-cell symmetric points specified in the parentheses
and the Madelung constant $\alpha_{\mathrm M}$ (the bulk energy
$\Ec_{\mathrm b}$ described by (\protect{\ref{Cq2}}) in units of
$-\kappa^2/a$). The positive ion is at the $(0,0,0)$ point. The repeating
potential values are indicated in symbols.} \label{Table2}
\begin{ruledtabular}
\begin{tabular}{lrrrrrrr}
Crystal&$U_{\mathrm{b1}}\:(0,\!0,\!0)$&$U_{\mathrm{b2}}\:(\frac{1}{4},\!%
\frac{1}{4},\!\frac{1}{4})\!$&$U_{\mathrm{b3}}\:({\scriptscriptstyle-}%
\frac{1}{4},\!\frac{1}{4},\!\frac{1}{4})\!\!\!\!$&$U_{\mathrm{b4}}\:(%
\frac{1}{2},\!0,\!0)$&$U_{\mathrm{b5}}\:(\frac{1}{2},\!\frac{1}{2},\!0)$&
$U_{\mathrm b6}\:(\frac{1}{2},\!\frac{1}{2},\!\frac{1}{2})$&
$\alpha_{\mathrm M}\qquad$\vspace{0.5mm}\\
\hline
&&&&&&&\vspace{-3.3mm}\\
CsCl&$-2.03536151$&$0\,\footnotemark[1]\qquad$&$U_{\mathrm{b2}}\lefteqn{
\footnotemark[1]}\qquad$&$0.48658923\lefteqn{\,$\footnotemark[1]$}$&$
\;-0.48658923\lefteqn{\,\footnotemark[1]}$&$\;\;2.03536151$&
$2.03536151$\\
NaCl&$-3.49512919$&$0\,\footnotemark[1]\qquad$&$U_{\mathrm{b2}}
\lefteqn{\footnotemark[1]}\qquad$&$3.49512919$&$U_{\mathrm{b1}}\qquad$
&$U_{\mathrm{b4}}\qquad$&$3.49512919$\\
ZnS&$-3.78292610$&$3.78292610$&$0.28779691\lefteqn{\,\footnotemark[1]}$&
$\;-0.28779691\lefteqn{\,\footnotemark[1]}$&$U_{\mathrm{b1}}\qquad$&
$U_{\mathrm{b4}}\lefteqn{\footnotemark[1]}\qquad$&$3.78292610$\\
Cu$_2$O&$-3.78292610$&$6.47653093$&$0.55497170\lefteqn{\,
\footnotemark[1]}$&$\;-0.28779691\lefteqn{\,\footnotemark[1]}$&
$U_{\mathrm{b1}}\qquad$&$U_{\mathrm{b4}}\lefteqn{\footnotemark[1]}\qquad$
&$\!10.25945703$\\
CaF$_2$&$-7.56585221$&$4.07072302$&$U_{\mathrm{b2}}\qquad$&
$-0.57559383\lefteqn{\,\footnotemark[1]}$&$U_{\mathrm{b1}}\qquad$&
$U_{\mathrm{b4}}\lefteqn{\footnotemark[1]}\qquad$&$\!11.63657523$\\
BiF$_3$&$-11.06098140$&$4.07072302$&$U_{\mathrm{b2}}\qquad$&
$2.91953536$&$U_{\mathrm{b1}}\qquad$&$U_{\mathrm{b4}}\qquad$&
$\!22.12196279$
\end{tabular}
\end{ruledtabular}
\footnotetext[1]{The corresponding lattice positions are interstitial.}%
\vspace{-2pt}
\end{table*}
infinity as long as the basic statements that $H$ is finite and that the
stationary limiting regime is provided by (\ref{Aq7}) are maintained.
As a result, the general concept that a rather arbitrary $\rho^{\mathrm
{ini}}(\rb)$ can be involved for describing a crystal \cite{Harr75,%
Goo69a,Oliv86,Herz79} is substantiated.

Note that our procedure of averaging the potential effect of the
bounding-box evolution resembles antecedent approaches \cite{Roy954,%
Cala76,Evje32,Gurn53,Dahl65,Hajj79,LeeC80}, which might be associated with
the averaging over a symmetric set of discrete points on the oscillatory
potential curve.

According to (\ref{Aq1}) and (\ref{Bq6}), the bulk potential field in a
crystal is as follows:
\begin{equation}\label{Bq7}
U_{\mathrm b}(\rb)=U_{\mathrm{Cd}}(\rb)+\mPhi_{\mathrm{top}} .
\end{equation}
Although relationship (\ref{Bq7}) is anticipated \cite{Harr70,Redl75}, the
topological treatment of $\mPhi_{\mathrm{top}}$ \cite{Khol01,Khol02}
differs from earlier interpretations \cite{Euwe75,SuCo95}. On combining
(\ref{Bq7}) and (\ref{Aq6}), for the mean bulk potential we obtain the
expected result $\bar{U}_{\mathrm b}=0$. Hence, $U_{\mathrm b}(\rb)$
defined by (\ref{Bq7}) is independent of $H$ and does not contain any
uniform component. The potential symmetry follows therefrom.
\vspace{-2pt}

\section{Discussion}\label{Sec4}
We may apply (\ref{Bq6}) and (\ref{Bq7}) to the rapidly convergent direct
Coulomb summation \cite{Fise92} for several cubic point-charge structures
with the lattice spacing $a$ and with charges measured in units of
$\kappa$, the modulus of a minimal point charge. These structures are
treated as decomposed into $n_L$ primitive lattices \cite{Tosi64,Hund35}
with the unit-cell charges compiled in Table~\ref{Table1}. According to
(\ref{Bq6}), for each structure in question we obtain
\begin{equation}\label{Cq1}
\mPhi_{\mathrm{top}}=\frac{\pi\kappa}{a}\times\!
\cases{\!4\bigl(s_6+s_7+9s_9\bigr)\!&$\lefteqn{\mathrm{CsCl}}$
\hspace{2.4em},\cr
\!4\bigl(s_3+4s_4+4s_6\bigr)\!&$\lefteqn{\mathrm{NaCl}}$
\hspace{2.4em},\cr
\!2\bigl(s_2+8s_4+11s_5+8s_7\bigr)\!&$\lefteqn{\mathrm{ZnS}}$
\hspace{2.4em},\cr
\!\bigl(s_2+16s_4+11s_5+9s_8\bigr)\!&$\lefteqn{\mathrm{Cu_2O}}$
\hspace{2.4em},\cr
\!4\bigl(s_2+4s_4+11s_5\bigr)\!&$\lefteqn{\mathrm{CaF_2}}$
\hspace{2.4em},\cr
\!4\bigl(s_2+s_3+4s_4+4s_6\bigr)\!&$\lefteqn{\mathrm{BiF_3}}$
\hspace{2.4em}}
\end{equation}
as the summary topological contribution of all $n_L$ sublattices. The
resulting numerical values calculated through (\ref{Aq1}) for each local
potential specified by (\ref{Bq7}) at some symmetric points are arranged
in Table~\ref{Table2}, where (\ref{Cq1}) retrieves the potential symmetry
\cite{Glas80,Tosi64,Goo69a,Leun82}.

Of course, dealing with an arbitrary $\rho^{\mathrm{ini}}(\rb)$, it is
reasonable to go over to $\rho(\rb)$ specified by $H=0$. Then the boundary
effects are formally excluded and the analysis is available in terms of
direct sums only. In this case $U_{\mathrm b}(\rb)=U_{\mathrm{Cd}}(\rb)$
that explains Evjen's approach \cite{Evje32} to NaCl, for example.
Moreover, the bulk Coulomb energy $\Ec_{\mathrm b}$ per unit cell can then
be confidently written in terms of (\ref{Aq1}) in a conventional manner
\cite{Bagn91,Dove83,Ewal21,Argy92,Emer23,Bert52,Leeu80,Wett80}:
\begin{eqnarray}\label{Cq2}
\Ec_{\mathrm b}&=&\frac{1}{2}\int_{V}d\rb\,\rho(\rb)\sump_i\int_V
\frac{\rho(\rb')\:d\rb'}{|\rb-\rb'-\Ri|}\nonumber\\
&=&\frac{1}{2}\int_{V}d\rb\,\rho(\rb)\,U_{\mathrm b}(\rb)=\frac{1}{2}
\int_{V^{\mathrm{ini}}}d\rb\,\rho^{\mathrm{ini}}(\rb)\,U_{\mathrm b}(\rb),
\qquad
\end{eqnarray}
in agreement with (\ref{Bq7}). The last relation in (\ref{Cq2}) is due to
the absolute character of the potential field $U_{\mathrm b}(\rb)$ and, in
general, implies that every charge distribution consistent with the
crystal structure may be incorporated there, if we are interested in the
value of $\Ec_{\mathrm b}$ at given $U_{\mathrm b}(\rb)$. In particular,
the Madelung constants in Table~\ref{Table2} are obtained with making use
of the potentials $U_{{\mathrm b}j}(\rb)$ on integer point charges, in
agreement with the corresponding precise data \cite{Glas80,Wadd59,Goo69a,%
Coke83,CraD87,Bhow88,Bens57,Saka58,John61,Sark88}. Bearing in mind
\cite{Khol01,Khol02} that the customary functional relation \cite{Bagn91,%
Gene87,Bert52}
\begin{equation}\label{Cq3}
\frac{\delta\Ec_{\mathrm b}}{\delta\rho^{\mathrm{ini}}(\rb)}
=U_{\mathrm b}(\rb)
\end{equation}
holds for (\ref{Cq2}) as a base for the self-consistent determination of
$\rho^{\mathrm{ini}}(\rb)$ \cite{Wadd59,Bagn91,Gene87,Dove83,Angy87,%
King93}, any difficulties \cite{Cole97,Harr75,Mart97,Leeu80,Oliv86,CraD87}
in defining $\Ec_{\mathrm b}$ are obviated via (\ref{Cq2}).

The topological effect of $\mPhi_{\mathrm{top}}$ is evident in the
hypothetical case of lattices composed of non-interacting neutral atoms,
which are to be spherical and non-overlapping for this purpose
\cite{Jack62}. In the simplest event \cite{Argy92,Bert52,Wett80,Herz79,%
Bert78,Hall81}, we consider such atoms as solid spheres of radii $r_0$,
charged uniformly, with neutralizing positive nuclear charges $Z$ at their
centres. One atom per unit cell is assumed. Then relations
(\ref{Aq2})--(\ref{Aq5}) hold by symmetry. According to (\ref{Aq5}) and
(\ref{Bq6}), we derive
\begin{equation}\label{Cq4}
\mPhi_{\mathrm{top}}=-\frac{2\pi Zr_0^2}{5v}\; .
\end{equation}
On the other hand, the sum in (\ref{Aq1}) is reduced to a sole term if
$\rb$ is inside an atom and is zero otherwise \cite{Jack62}. The
calculation of the mean value $\bar{U}_{\mathrm{Cd}}$ is then
straightforward and we get
\begin{equation}\label{Cq5}
\bar{U}_{\mathrm{Cd}}=\Bigl(\frac{4\pi r_0^3}{3v}\Bigr)\bar{U}
\quad\mbox{at}\quad\bar{U}=\frac{3Z}{10r_0}\; ,
\end{equation}
where $\bar{U}$ is the mean intratomic potential, the factor in the
parentheses renormalizes $\bar{U}$ per $v$. According to (\ref{Bq7}),
(\ref{Cq4}) and (\ref{Cq5}), we reach $\bar{U}_{\mathrm b}=0$ again.
However, as a centre of gravity \cite{Bagn91,Spac81,Euwe75,Herz85},
$\mPhi_{\mathrm{top}}$ now connects intratomic potentials which dwell as
formally independent in the case at hand. It is worth noting that this
effect of (\ref{Cq4}) is absent in low-dimensional lattices \cite{Harr75,%
Emer55,Coke83,IhmC80,Hall79} where $\mPhi_{\mathrm{top}}=0$. But such a
negative shift due to delocalization of electrons in atoms and ions is a
general effect which, apart from the effects of ionic overlap
\cite{Harr75,Dove83,Birm55}, modifies the three-dimensional point-charge
lattice potentials. According to estimate (\ref{Cq4}), this shift is
compared to potentials in Table~\ref{Table2} and may account for trapping
small atoms and ions at interstices \cite{Harr70,Macd82,Isoy90,Lowt95,%
Bona98} as well as for the asymmetry of vacancy concentrations
\cite{Goo69b,Bogu95,Oate95,Piqu97,Gorc99,Kho02b}.

Note that every artificial delocalization of charges \cite{Fuch94,%
Birm55,Ewal21,Argy92,Bert52,Sugi84,Luty95,HeyS81,LeeC80,SuCo95,Herz79}
always produces a certain potential shift of the discussed nature, with
including uniform charge distributions \cite{Bagn91,Ewal21,Heye81,Klei81,%
Wett80,Keef94,IhmC80,Hall79}. For instance, according to (\ref{Aq5}) and
(\ref{Bq6}), in the case of simple cubic lattices a uniform charge
backgroung with the total charge $Q$ per unit cell gives rise to
$\mPhi_{\mathrm{top}}=\pi Q/6a$. This term compensates the ordinary
background effect \cite{Keef94}. However, such contributions cancel out in
the superposition if, for example, two identical sublattices composed of
unlike charges, but with common parameters of charge delocalization are
studied \cite{Ewal21}.

\section{Conclusion}\label{Sec5}
In summary, a special mode of summation associated with invariant periodic
boundary conditions must be included in the definition of Coulomb lattice
series so as to describe bulk properties of crystals. As a result,
absolute electrostatic potentials arise as a unique solution of interest.
Zero mean bulk potential ensuring this uniqueness within every summation
scheme devised to date is substantiated rigorously. Based on absolute
bulk potentials, the invariant treatment of the bulk Coulomb energy is
proposed. The limiting cases estimated highlight the potential symmetry as
a peculiar feature of the three-dimensional potential map.

\begin{acknowledgments}
I am grateful to Professor V.~L. Ginzburg, Professor E.~G. Maksimov and
Professor V.~G. Vaks for their encouragement of this work. I thank Miss
J.~E. Kholopova for her help in searching reference sources. I am indebted
to architect B.~V. Kholopov for his support.
\end{acknowledgments}

\end{document}